\documentclass{article}
\begin{document}
\noindent{\Large\bf Nearly integrable nonlinear equations on
the half line}

\bigskip
\noindent 
E V Doktorov\footnote{E-mail: doktorov@dragon.bas-net.by}
and V S Shchesnovich\footnote{E-mail: valery@maths.uct.ac.za}
\\

\noindent${}^1$B I Stepanov Institute of Physics, 68 F
Skaryna Ave, 220072 Minsk, Belarus\\
${}^{2}$Department of Mathematics and Applied
Mathematics, University of Cape Town, University Private Bag,
Rondebosch 7701, South Africa\footnote{On leave from the Division for 
Optical Problems in Information Technologies, National Academy of 
Sciences of Belarus, Zhodinskaya St. 1/2, 220141 Minsk, Belarus}


\begin{abstract}
We employ the Riemann-Hilbert problem for solution of the
initial-boundary value problems for nearly integrable equations on the
half line which have important applications in physics.
The detailed derivation of the integrable and perturbation-induced
evolutions of the spectral data
are given for the two most important models, the self-induced
transparency and the stimulated Raman scattering. In particular,
we prove that dephasing stabilizes the spike phenomenon in the
stimulated Raman scattering.
\end{abstract}

\section{Introduction}\label{sec:In}

Nonlinear integrable models of resonant interaction of radiation
with matter possess a number of features which distinguish them
from a variety of integrable systems. First of all, these
equations determine initial-boundary-value
problems, because, along with an initial condition, some boundary conditions
are necessary to properly formulate the problem. Secondly, they have
singular dispersion relations~\cite{1,2}. As examples of such integrable
equations we mention  the Maxwell-Bloch equations of self-induced
transparency~\cite{3}, equations of stimulated Raman scattering~\cite{4},
the Karpman-Kaup equations~\cite{5,6}, the equations of the
interaction of the electrostatic high-frequency wave with the
ion-acoustic wave~\cite{7}, and the equations of the resonant
interaction between optical solitons and impurity atoms in
fibres~\cite{8}.

There exists a well-developed formalism for solution of the
integrable equations with the singular dispersion relations when
the boundary conditions are asymptotic, i. e. they are posed at
the plus or/and minus infinity \cite{9,10,11}. Along with the
fact that the semi-infinite interval, when one of the boundaries
is placed at a finite point, e. g. zero, is physically more
appropriate, the formalism of the inverse scattering has to be
substantially modified as compared with the  whole line
case~\cite{12,13}.

Recently significant progress has been achieved in solving
the nonlinear evolution equations with singular dispersion relations
posed on the semi-infinite~\cite{14,15} and finite~\cite{16,17}
intervals. Also it was realised that the Riemann-Hilbert (RH)
problem is a natural setting for  analysis of the above
nonlinear models. The data uniquely characterising the solution to the RH
problem serve as parameters of the solution to these nonlinear equations.

All the above-mentioned papers deal with the {\it integrable}
evolutions. On the other hand, in the experimental observations
one has various additional (frequently small enough) effects
which destroy the integrability property but must be accounted for
complete explanation of the experimental results. For instance, the
phenomenon of the spike formation in the pump depletion zone in
the stimulated Raman scattering~\cite{18} is directly related
with the dephasing effect in the medium~\cite{19}.
Hence, a problem arises to describe evolution of the
RH (or spectral) data in the presence of small perturbations, i. e.,
to develop a perturbation theory on the half line.

In the present paper we extend the perturbation theory developed for the
whole line \cite{20,21}, to the half line.
We take the equations of the self-induced transparency (SIT) and stimulated
Raman scattering (SRS) posed on the half line as the examples.
In both cases the dephasing effect, which destroys the integrability 
of the models, is
considered as the perturbation. We show that for the half line, as for the
whole line,  construction of the perturbation theory amounts to introduction
of an additional matrix functional, which we call the evolution  functional.
Moreover, it turns out that the evolution functional has the explicit 
expression very much
similar to that in the case of the whole line (see, for instance 
\cite{21}).

The plan of the paper is the following.
In section 2 we briefly remind the formulation of the initial-boundary
value problems for the SIT and SRS equations. In sections 3 and 4
we present necessary results on the direct and, respectively, 
inverse spectral problems
for the half line. In particular,  we demostrate the advantage of
dealing with the analytic functions with constant determinant for 
derivation of the evolution equations for the spectral data in 
the case of the half  line.
Our main section is section 5, where  the perturbation-induced evolution
equations for the RH data are derived.

This paper is dedicated to Professor Pierre Sabatier whose versatile 
activities have determined in many respects the contemporary status 
of inverse methods in science and technology,  
in honor of his 65th birthday.

\bigskip

\section{Nonlinear models with singular dispersion relations}

The SIT phenomenon consists in a soliton-like $2\pi$-pulse
propagation in a medium without energy loss and is described by
the Maxwell-Bloch equations~\cite{3}
\begin{equation}
{\cal E}_x=<\rho>,\quad \rho_t+2i\lambda\rho+\Gamma\rho=N{\cal
E},\quad N_t=-\frac{1}{2}\left({\cal E}\bar\rho+\bar{\cal
E}\rho\right).
\label{1}\end{equation}
Here ${\cal E}(x,t)$ is the envelope of the light-pulse electric field 
with the carrier frequency close to the particular transition 
frequency of an atom,
\[
M=\left(\begin{array}{cc}-N&\rho\\ \bar\rho&N\end{array}\right)
\]
is the density matrix of the atomic ensemble (medium) with
$\rho(x,t,\lambda)$ being induced polarisability and
$N(x,t,\lambda)$ being normalised population difference between
the chosen atomic levels. We have $\mbox{det}M=-1$ and for atoms
initially on the lower level $N(t=0)=-1$ and $M(t=0)=\sigma_3$.
$\lambda$ is the normalised frequency mismatch between the carrier
frequency and atomic transition frequency, $\Gamma$ is the
dephasing factor. We think of the value of $\Gamma$ as being
small, $\Gamma\ll 1$, in order to justify the applicability of the
perturbation theory. The coordinate $x$ is the resonant medium
extension and $t$ is time. The symbol $<\dots>$ means the
averaging over the frequency mismatch with the normalised
distribution function $g(\lambda)$:
$$
<\rho>=\int_{-\infty}^\infty{\rm
d}\lambda\rho(x,t,\lambda)g(\lambda), \qquad
\int_{-\infty}^\infty{\rm d}\lambda g(\lambda)=1.
$$
As regards the SRS equations, we follow the approach by Leon and
Mikhailov~\cite{14}:
\begin{equation}
u_t+\Gamma u=-g\int_{-\infty}^\infty{\rm d}ka_1\bar
a_2e^{-2ikx},\quad a_{1x}=ua_2e^{2ikx},\quad a_{2x}=-\bar
ua_1e^{-2ikx}. 
\label{2}\end{equation} 
Here $a_1(x,t,k)$ and
$a_2(x,t,k)$ are the electric field envelopes of the pump and
Stokes pulses, respectively, the parameter $k$ describes a
spectral extension of the input pulses, $u(x,t)$ is related with
the polarisability of a Raman-active medium, $g$ is a coupling
constant. It should be noted that the parameter $k$ can be
connected with the group-velocity dispersion of the pulses in the
medium~\cite{14}.

Both nonlinear systems are considered in the quadrant $0\le
x<\infty$, $0\le t<\infty$ with the initial and boundary values:
\begin{eqnarray}
{\rm a)}\; {\rm  SIT:}& & \quad{\cal E}(0,t)={\cal E}_0(t),\quad
M(x,0,\lambda)\equiv M_0=\left(\begin{array}{cc}-N_0&\rho_0\\
\bar\rho_0&N_0\end{array}\right),
\quad {\cal E} \to 0\quad {\rm as} \quad t\to\infty;
\label{3}\\[.5cm]
{\rm b)}\;{\rm SRS:}& & \quad u(x,0)=0,\quad a_1(0,t,k)=I_1(t,k),\quad
a_2(0,t,k)=I_2(t,k), \label{4}\end{eqnarray}
Note that for the SIT equations we have an
initial-boundary-value problem, while for the SRS equations we consider
a pure boundary-value problem.

For $\Gamma=0$ both systems (1) and (2) are integrable and admit
the Lax representation~\cite{3,22}. 
In both cases  the dispersion relation  is
non-analytic (singular) function on the 
complex plane of the spectral parameter.

\bigskip

\section{Direct spectral problem}

Because both the SIT and SRS equations are solved by means of one
and the same Zakharov-Shabat spectral problem~\cite{23}, we will
analyze below the direct spectral problem for both systems
together (omitting below the explicit dependence on the other coordinate).
The spectral problem is written as
\begin{equation}
\chi_\xi=-ik[\sigma_3,\chi]+Q\chi. \label{5}\end{equation}
Here    $\chi$ is the $2\times2$ matrix function,
\[
Q=\left(\begin{array}{cc}0&q\\-\bar q&0\end{array}\right)
\]
is a potential with ${\cal E}=2q$ for the SIT and $u=q$ for the SRS. The
coordinate $\xi$ stands for $t$ in the case of the SIT and for $x$ in
the SRS case, $k$ is a spectral parameter coinciding with the
mismatch parameter for SRS. It should be stressed that the
physical status of the potential $q$ is different for the SIT and
SRS. Whereas for the SIT the potential $q$ (or ${\cal E}$) is the
main experimentally measured quantity, for the SRS the potential $u$
could be reconstructed only indirectly from the measured
intensities $|a_1|^2$ and $|a_2|^2$.

Let us define the matrix Jost solutions $J_\pm(\xi,k)$ of the spectral
equation (5) by means of the following conditions:
\begin{equation}
J_-(0,k) = I,\quad J_+(\xi,k)\to I,\quad {\xi\to\infty},
\label{jost}\end{equation}
where $I$ is the identity matrix.  The scattering matrix is
defined in the usual way:
\begin{equation}
S=E^{-1}J_+^{-1}J_-E,\qquad E(\xi,k)=e^{-ik\xi\sigma_3}, \qquad
{\rm Im}k=0
\label{6}\end{equation}
and has the structure
\begin{equation}
S=\left(\begin{array}{cc}a&-\bar b\\b&\bar a\end{array}\right),
\qquad a\bar{a}+b\bar{b}=1.
\label{S}\end{equation}
The scattering matrix is connected with the boundary value of
$J_+$, $S(k)=J_+^{-1}(0,k)$. Note that the potential is anti-Hermitian:
$Q^\dag=-Q$. Hence, we have the  involutions (${\rm Im}k=0$):
\begin{equation}
J_\pm^\dag(\xi,k)=J^{-1}_\pm(\xi,k),\quad
S^{\dag}(k)=S^{-1}(k).
\label{invol}\end{equation}
The spectral equation (\ref{5}) can be represented in the integral form:
\begin{equation}
J(\xi,k)=E(\xi)\left[J_0+\int_{\xi_0}^\xi{\rm
d}\xi'E^{-1}(\xi')Q(\xi')J(\xi',k)E(\xi')\right]E^{-1}(\xi),
\label{7}\end{equation} where $J_0=J(\xi=\xi_0)$. The Jost
solution $J_-$ ($J_+$) corresponds to the choice $\xi_0=0$
($\xi_0=\infty$). It follows from (\ref{7}) that the matrix functions
$(J_-^{(1)},J_+^{(2)})$  and $(J_+^{(1)},J_-^{(2)})$ are solutions
to equation (\ref{5}) and moreover holomorphic in $k$ in the upper and lower half-planes,
$\pm{\rm Im}k\ge0$, respectively. Here $J^{(n)}$ stands for the corresponding
column vector. By noticing that a solution to (\ref{5}) transforms to another
solution if multiplied on the right by a diagonal
$\xi$-independent matrix and that ${\rm det}(J_-^{(1)},J_+^{(2)})=a$ and
${\rm det}(J_+^{(1)},J_-^{(2)})=\bar{a}$ we introduce the  solutions to
(\ref{5}) having unit determinant:
\begin{equation}
\Psi_+=\left(J_-^{(1)},a^{-1}J_+^{(2)}\right),\qquad
\Psi_-=\left(\bar a^{-1}J_+^{(1)},J_-^{(2)}\right).
\label{11}\end{equation}
By definition the functions $\Psi_+(\xi,k)$ and $\Psi_-(\xi,k)$ are analytic in the upper and
lower half-planes of the complex $k$-plane, respectively.
On the real line ${\rm Im}k=0$ they can be given in terms of
the matrix Jost solutions and entries of the scattering matrix:
\begin{equation}
\Psi_+=J_+EG_+E^{-1}=J_-EG_-E^{-1},\quad
\Psi_-=J_+EH^{-1}_+E^{-1}=J_-EH^{-1}_-E^{-1}.
\label{10}\end{equation}
Here
$$
G_+=\left(\begin{array}{cc}a&0\\b&1/a\end{array}\right),\quad
G_-=\left(\begin{array}{cc}1&\bar
b/a\\0&1\end{array}\right),
$$
$$
H_+=\left(\begin{array}{cc}\bar a&\bar b\\0&1/\bar
a\end{array}\right),\quad H_-=\left(\begin{array}{cc}1&0\\b/\bar
a&1\end{array}\right).
$$
As follows from (\ref{invol}) and (\ref{10}),
the matrix functions $\Psi_\pm(\xi,k)$ satisfy the involution:
\begin{equation}
\Psi^\dag_+(k)=\Psi^{-1}_-(\bar{k}). \label{invol2}\end{equation}
The representation (\ref{10}) involves the factorizations of the
scattering matrix {\it with division}: $S=G_+G_-^{-1}$ and
$S=H_+^{-1}H_-$. The advantage of using such factorizations for
the half line stems from the fact that both $G_-$ and $H_-$
contain a {\it single} quantity: $\bar{b}/a$ and $b/\bar{a}$,
respectively. On the other hand they provide the values of
$\Psi_\pm$ at the boundary: $\Psi_+(0,k)=G_-$ and
$\Psi_-(0,k)=H^{-1}_-$. In view of this, it is convenient to
introduce the following notations:
\begin{equation}
\beta(k)=\frac{\bar{b}(k)}{a(k)},\quad
\alpha(k)=\frac{1}{a(k)}.
\label{alphabeta}\end{equation}
The functions $\beta(k)$ and $\alpha(k)$ are usually referred
to as the reflection and transmission coefficients, respectively.
Then the boundary values of the solutions $\Psi_\pm(\xi,k)$ read
\begin{equation}
\Psi_+(\xi=0)=\left(\begin{array}{cc}1&\beta\\0&1\end{array}\right),
\qquad
\Psi_-(\xi=0)=\left(\begin{array}{cc}1&0\\-\bar\beta&1\end{array}\right).
\label{12}\end{equation}
For completeness, let us present also the asymptotic values of $\Psi_\pm(\xi,k)$ as
$\xi\to\infty$:
\begin{equation}
\Psi_+(\xi)\to\left(\begin{array}{cc}1/\alpha&0\\{\bar\beta}/{\bar\alpha}\,
e^{2ik\xi}&\alpha\end{array}\right),\qquad
\Psi_-(\xi)\to\left(\begin{array}{cc}\bar \alpha&{-\beta}/{\alpha}\,
e^{-2ik\xi}\\0&1/{\bar\alpha}\end{array}\right).
\label{asymp}\end{equation}

Now, let us consider in more detail
the analyticity properties and the $k$-asymptotics
of the matrix functions $\Psi_\pm$.
These properties are easy to deduce from the integral
equations for these functions:
\[
\left(\begin{array}{c}\Psi_{+11}\\\Psi_{+21}\end{array}\right)=\left(\begin{array}{c}
1\\0\end{array}\right)+\left(\begin{array}{c}\int_0^\xi{\rm
d}\xi'q\Psi_{+21}\\ -\int_0^\xi{\rm d}\xi'e^{2ik(\xi-\xi')}\bar
q\Psi_{+11}\end{array}\right),
\]
\[
\left(\begin{array}{c}\Psi_{+12}\\\Psi_{+22}\end{array}\right)=\left(\begin{array}{c}
0\\1\end{array}\right)-\left(\begin{array}{c}\int_\xi^\infty{\rm
d}\xi'e^{-2ik(\xi-\xi')}q\Psi_{+22}\\\int_0^\xi{\rm d}\xi'\bar
q\Psi_{+12}\end{array}\right),
\]
\[
\left(\begin{array}{c}\Psi_{-11}\\\Psi_{-21}\end{array}\right)=\left(\begin{array}{c}
1\\0\end{array}\right)+\left(\begin{array}{c}\int_0^\xi{\rm
d}\xi'q\Psi_{-21}\\ \int_\xi^\infty{\rm
d}\xi'e^{2ik(\xi-\xi')}\bar q\Psi_{-11}\end{array}\right),
\]
\[
\left(\begin{array}{c}\Psi_{-12}\\\Psi_{-22}\end{array}\right)
=\left(\begin{array}{c}
0\\1\end{array}\right)+\left(\begin{array}{c}\int_0^\xi{\rm
d}\xi'e^{-2ik(\xi-\xi')}q\Psi_{-22}\\-\int_0^\xi{\rm d}\xi'\bar
q\Psi_{-12}\end{array}\right).
\]
Evidently the first column $\Psi_+^{(1)}$
and the second column $\Psi_-^{(2)}$ are entire functions of $k$,
while $\Psi_+^{(2)}$ and $\Psi_-^{(1)}$ are meromorphic
functions in the upper and lower  half-planes of the complex $k$-plane,
respectively (see also Ref.~\cite{14}).
By application of (\ref{12}) we conclude that the reflection
coefficient $\beta(k)$ is meromorphic in the upper half of the
complex plane. Moreover, as $\beta/\alpha=\bar{b}$, by using the
identity $J_{+12}(0,k)= \bar{b}(k)$ (see (\ref{6}) and (\ref{S}))
we conclude that the quotient $\beta(k)/\alpha(k)$ is holomorphic in the
upper half plane. Equivalently, $a(k)$ and $\bar{b}(k)$ are holomorphic
in the upper half of the comlex $k$-plane.
Note that the poles of $\beta(k)$ coincide with zeros of $a(k)$.

From the above integral equations it is seen that
$\Psi_\pm \to I$ as $k\to\infty$. Integration by parts in the
integrals containing $\exp(\pm2ik\xi')$ gives the {\it generalized} asymptotic
expansions:
\[
\left(\begin{array}{c}\Psi_{+11}\\\Psi_{+21}\end{array}\right)=\left(\begin{array}{c}
1\\0\end{array}\right)+\frac{1}{2ik}\left(\begin{array}{c}\int_0^\xi{\rm
d}\xi'|q(\xi')|^2-\bar q(0)\int_0^\xi{\rm
d}\xi'e^{2ik\xi'}q(\xi')\\\bar q(\xi)-e^{2ik\xi}\bar
q(0)\end{array}\right)+{\cal O}\left(\frac{1}{k^2}\right),
\]
\[
\left(\begin{array}{c}\Psi_{+12}\\\Psi_{+22}\end{array}\right)=\left(\begin{array}{c}
0\\1\end{array}\right)+\frac{1}{2ik}\left(\begin{array}{c}q(\xi)\\-\int_0^\xi{\rm
d}\xi'|q(\xi')|^2\end{array}\right)+{\cal
O}\left(\frac{1}{k^2}\right),
\]
\[
\left(\begin{array}{c}\Psi_{-11}\\\Psi_{-21}\end{array}\right)=\left(\begin{array}{c}
1\\0\end{array}\right)+\frac{1}{2ik}\left(\begin{array}{c}\int_0^\xi{\rm
d}\xi'|q(\xi')|^2\\\bar q(\xi)\end{array}\right)+{\cal
O}\left(\frac{1}{k^2}\right),
\]
\[
\left(\begin{array}{c}\Psi_{-12}\\\Psi_{-22}\end{array}\right)=\left(\begin{array}{c}
0\\1\end{array}\right)+\frac{1}{2ik}\left(\begin{array}{c}q(\xi)-e^{-2ik\xi}q(0)\\
-\int_0^\xi{\rm d}\xi'|q(\xi')|^2+q(0)\int_0^\xi{\rm
d}\xi'e^{-2ik\xi'}\bar q(\xi')\end{array}\right)+{\cal
O}\left(\frac{1}{k^2}\right).
\]
These formulae give triangular matrices for $\Psi_\pm(0,k)$
in agreement with the the boundary values (\ref{12}). For ${\rm Im}k=0$ and $\xi\ne0$
$\Psi_\pm(\xi,k)$ do not have asymptotic expansions in the {\it usual} sense
due to the presence of the terms $ \exp(2ik\xi)\bar{q}(0)$ and $\exp(-2ik\xi)q(0)$.

Now we are in a position to express the potential $Q$ through the
matrix elements of $\Psi_\pm$:
\begin{equation}
q(\xi)=2i\lim_{k\to\infty}k\Psi_{+12}(\xi,k),
\quad
\bar q(\xi)=2i\lim_{k\to\infty}k\Psi_{-21}(\xi,k).
\label{13}\end{equation}
Alternatively, one can use the following formulae:
\[
q(\xi)=2i\lim_{k\to\infty}k\left(\Psi_{-12}(\xi,k)+e^{-2ik\xi}\beta(k)\right),
\]
\[
\bar q(\xi)=2i\lim_{k\to\infty}k\left(\Psi_{+21}(\xi,k) - e^{2ik\xi}\bar\beta(k)\right).
\]
Here we have taken into account  (\ref{13})  for $\xi=0$ with
 $\Psi_{+12}(0,k)=\beta(k)$ and $\Psi_{-21}(0,k) = -\bar{\beta}(k)$ from (\ref{12}).

The reconstruction of the potential requires knowledge of the functions $\Psi_\pm$.
The latter can be obtained via solution of the RH problem, which we consider in the next section.
\bigskip

\section{The Riemann-Hilbert problem}

We have from (\ref{10})
\begin{equation}
 \Psi_+=\Psi_- EGE^{-1}, \qquad {\rm Im}k=0,
\label{19}\end{equation}
where
\begin{equation}
G(k)=H_-G_-=H_+G_+=\left(\begin{array}{cc}1&\beta\\\bar \beta&
1+|\beta|^2\end{array}\right).
\label{20}\end{equation}
This is the matrix RH problem, i. e., the problem
of analytic factorization a nondegenerate matrix $G(k)$, given on the
real line ${\rm Im}k=0$, into a product of two
matrices analytic  in the upper and lower half-planes,
respectively. The above RH problem has the canonical normalization condition:
\begin{equation}
\Psi_\pm\to I \qquad {\rm as} \quad k\to\infty.
\label{21}\end{equation}
To completely characterize the solution to the RH
problem (\ref{19})-(\ref{21}) one needs to specify the
spectral data. First, it is the value of $\beta(k)$ on the real line
${\rm Im}k=0$, which enters $G(k)$ (\ref{20})
and represents the continuous part of the spectral data.
Second, the discrete spectral data should be given.
The latter include zeros, $k_j$, ${\rm
Im}k_j>0$, of the function $\beta^{-1}(k)$ [$\beta^{-1}(k_j)=0$]
and the transition coefficients $c_j$. The transition coefficients are given
by the residues of $\Psi_+$ or $\Psi_-$  at the points $k_j$ and $\bar k_j$,
respectively. Using the identity ${\rm det}\Psi_\pm(\xi,k)=1$, the involution
(\ref{invol2})  and formula (\ref{11})
it is not difficult to obtain (see also \cite{14})
\begin{equation}
{\rm Res}\left[\Psi_+^{(2)}(k),k_j\right]=c_j\Psi_+^{(1)}(k_j),
\quad {\rm Res}\left[\Psi_-^{(1)}(k),\bar k_j\right]=-\bar
c_j\Psi_-^{(2)}(\bar k_j). \label{22}\end{equation}
The $\xi$-dependence of $c_j$ follows from the equation
\[
\partial_\xi\left[{\rm
Res}\Psi_+^{(2)},k_j\right]=-ik\left(\sigma_3+I\right)\left[{\rm
Res}\Psi_+^{(2)},k_j\right]+Q\left[{\rm
Res}\Psi_+^{(2)},k_j\right].
\]
Substituting here (\ref{22}), we get
$c_j(\xi)=c_j^{(0)}e^{-2ik_j\xi}$.

The last step towards the complete description of the RH data
involves the analysis of their temporal behavior. Here we analyse in detail
the two models of section 2, the SIT and SRS equations.
These models are integrable and hence represent the compatibility
condition of a linear system, i. e., the Lax pair.
The first Lax equation is the spectral equation (\ref{3}).
The time evolution is determined by the second Lax equation:
\begin{equation}
\partial_\tau\Psi_{\pm}  =\Psi_\pm  E \Omega_\pm E^{-1}
+ V_\pm \Psi_\pm .
\label{second}\end{equation}
Here $\tau=x$ for the SIT and $\tau=t$ for the SRS models.
The matrices $V_\pm(\xi,\tau,k)$ are the limitimg values,
as $k\to{\rm Re}k\pm i0$, of a piece-wise
holomorphic function $V(\xi,\tau,k)$.
The choice of $ V(\xi,\tau,k)$ depends on the model.
In the same way, the dispersion relation $\Omega(k,\tau)$
is also a piece-wise holomorphic fuction with the limiting
values $\Omega_\pm(k,\tau)$ defined for real  $k$.
(Hence the name ``singular", i. e., non-analytic  dispersion relation).
The matrices $V$ and $\Omega$  satisfy the
involutions
\[
V^\dag_+(\bar{k})=-V_-(k),\quad
\Omega^\dag_+(\bar{k})=-\Omega_-(k),
\]
which follow from those of $\Psi(k)$ (\ref{invol2}).
Note that the compatibility of the Lax pair (\ref{3}) and (\ref{second}) is
guaranteed if $V$ is chosen in such a way that the jump matrix
$\Delta V(\xi,\tau,k)= V_+(\xi,\tau,k) - V_-(\xi,\tau,k)$ satisfies
\begin{equation}
\partial_\xi \Delta V(\xi,\tau,k)  = [-ik\sigma_3 + Q,\Delta V(\xi,\tau,k)],
\label{genequ} \end{equation} while  $\Omega(k,\tau)$ is {\it
still arbitrary}. It means that the dispersion relation  is {\it
not} defined by the model. In fact, as it was realized in
Ref.~\cite{11}, it is specified by the imposed boundary
conditions (see also \cite{14}).  Let us first obtain $V$ for the
models we consider. It is convenient to represent a $V$
satisfying (\ref{genequ}) in the form
\begin{equation}
V(\xi,\tau,k)= \frac{1}{2\pi i}\int_{-\infty}^\infty\frac{{\rm
d}\lambda}{\lambda-k}J_-(\xi,\tau,\lambda)E(\xi,\lambda)\Delta V(0,\tau,\lambda)
E^{-1}(\xi,\lambda)J_-^{-1}(\xi,\tau,\lambda),
\label{general}\end{equation}
Formula (\ref{general}) trivially transforms  in the SIT case by letting
\begin{equation}
\Delta V(\xi=0) = - \frac{\pi g(k)}{2} M_0,
\label{SIT}\end{equation} while for the SRS equations we have
\begin{equation}
\left(\begin{array}{c}a_1 \\ a_2 e^{2ik\xi}\end{array}\right)
= I_1 J^{(1)}_- + I_2 J^{(2)}_- e^{2ik\xi}\equiv I_1 \Psi_+^{(1)} + I_2 \Psi_-^{(2)}e^{2ik\xi},
\label{relat}\end{equation}
thereby
\begin{equation}
 \Delta V(\xi=0) = \frac{\pi{g}}{2}\left(\begin{array}
{cc}|I_1|^2-|I_2|^2, &
2 {I}_1 \bar{I}_2 \\ 2\bar{I}_1{I}_2
& |I_2|^2-|I_1|^2,
\end{array}\right).
\label{SRS}\end{equation}
Note that in both cases ${\rm tr}V= {\rm tr}\Delta V=0$.

To derive evolution equations for the spectral data and the
dispersion relation $\Omega$ we recall that the analyticity
properties of $\Psi_\pm(k)$ depend  on the conditions
(\ref{jost}) defining the Jost solutions to the spectral equation
(\ref{5}). Hence, to have analyticity of $\Psi_\pm(k)$ for
arbitrary $\tau$ one must demand that those conditions hold for
all $\tau$. Equivalently, we can deal with the boundary and
asymptotic values of $\Psi_\pm(\xi,\tau,k)$, which must satisfy
(\ref{12}) and (\ref{asymp}) for arbitrary $\tau$. Substitution of
formulae (\ref{12}) and (\ref{asymp}) into the second Lax
equation (\ref{second}) supplies one with the evolution of the
spectral data $\beta$ and $\alpha$ together with the dispersion
relation $\Omega$. After somewhat lengthy but straightforward
calculations (in computing the limits of $V_\pm$ as $\xi\to
\infty$ one uses the identity: ${V.p.}\{\exp(\pm2ik\xi)/k\} \to
\pm i\pi \delta(k)$) we obtain (see also Ref.~\cite{14})
\begin{equation}
\beta_\tau = -V_{+21}^{(0)}\beta^2+2V^{(0)}_{+11}\beta+V^{(0)}_{+12},
\label{29}\end{equation}
\begin{equation}
\alpha_\tau = \left( (S\Delta V^{(0)} S^{-1})_{22} - 
\beta V^{(0)}_{+21} - V^{(0)}_{+22}
\right)\alpha.
\label{foralpha}\end{equation}
Here $V_{+ij}^{(0)}\equiv V_{+ij}(\xi=0)$.
For completeness we give also the dispersion relation:
\[
\Omega_+=\left(\begin{array}{cc}\beta
V_{+21}^{(0)}-V_{+11}^{(0)}&0\\-V_{+21}^{(0)}&-\beta
V_{+21}^{(0)}-V_{+22}^{(0)}\end{array}\right),
\]
\[
\Omega_-=\left(\begin{array}{cc}\bar\beta
V_{-12}^{(0)}-V_{-11}^{(0)}&-V_{-12}^{(0)}\\0&-\bar\beta
V_{-12}^{(0)}-V_{-22}^{(0)}\end{array}\right).
\]

Here we note that the corresponding evolution equations for the entries
($a$ and $b$) of the scattering matrix are highly nonlinear
coupled equations;
while  for the quotient  $\beta=\bar{b}/a$ one gets a
simple Riccati equation.
This is the advantage of using the matrices $\Psi_\pm$
defined by (\ref{11}).
The relevance of the Riccati equation for equations with the
singular dispersion relations was noted previously (see, for
example,~\cite{25} for SIT and~\cite{14, 26} for the SRS).

The time evolution of the discrete RH data is found as follows.
First, let us find the time dependence of $c_j$. It follows 
from (\ref{22}) taken at $\xi=0$ that
\[
c_j^{(0)}(\tau)={\rm Res}[\beta(k,\tau),k_j],
\]
thereby
\begin{equation}
c_j(\xi,\tau)=e^{-2ik_j\xi}{\rm Res}[\beta(k,\tau),k_j].
\label{32}\end{equation}
Evolution of the poles of $\beta$ is derived in the following way.
Inasmuch as $\beta^{-1}(k_j,\tau)=0$ for all $\tau$, we have
\begin{equation}
\left. \frac{{\rm d}k_j}{{\rm d}\tau}= -\frac{(\partial/\partial\tau)\beta^{-1}(k)}{(\partial/\partial
k)\beta^{-1}(k)}\right|_{k=k_j}
\label{kj}\end{equation}
which, in view of (\ref{29}) and the identity
$({\partial} /{\partial k})\beta^{-1}(k_j)=1/c_j^{(0)}$, gives
\begin{equation}
\frac{{\rm d}k_j}{{\rm d}\tau}= -\frac{V_{+21}^{(0)}(k_j)}{\partial_k\beta^{-1}(k_j,\tau)}
= - V_{+21}^{(0)}(k_j) c^{(0)}_j.
\label{31}\end{equation}
We remind that
\[
{\rm a)}\; {\rm SIT:}\qquad V_{+21}^{(0)}(x,k_j)=\frac{i}{4}\int_{-\infty}^\infty
\frac{{\rm d}\lambda}{\lambda-k_j}g(\lambda)  \bar {\rho}^{(0)}(x,\lambda)
\]
\[
\;{\rm b)}\; {\rm SRS:}\qquad  V_{+21}^{(0)}(t,k_j)=\frac{g}{2i}\int_{-\infty}^\infty
\frac{{\rm d}\lambda}{\lambda-k_j}\bar I_1(t,\lambda)I_2(t,\lambda).
\]
Hence, $k_j={\rm const}$ for the
``causal" solutions (in Zakharov's terminology~\cite{27}) for
the SIT or in the absence of the Stokes pulse for the SRS.

In this paper we are interested primarily in description of the
effect of a  perturbation on the evolution of the spectral data
rather than characterization of specific solutions to the
integrable nonlinear equations. Therefore, we proceed to the
derivation of the perturbation-induced evolution of the spectral
data. This is done in the next section.

\bigskip

\section{Perturbation-induced evolution of RH  data}

Let us attribute the ``variational" derivatives to the
perturation-induced evolution.
For instance,  we introduce the perturbation matrix
$R$ as the variation of the potential,
\begin{equation}
R=\frac{\delta Q}{\delta \tau}=\left(\begin{array}{cc}0&\delta
q/\delta\tau\\-\delta\bar q/\delta\tau&0\end{array}\right)\equiv
\left(\begin{array}{cc}0&r\\-\bar r&0\end{array}\right).
\label{33}\end{equation}
A variation of $Q$ leads to the corresponding
variations $\delta J_\pm$ of the Jost solutions. We have
$$
\left(J_\pm^{-1}\delta
J_\pm\right)_\xi=-ik\left[\sigma_3,J_\pm^{-1}\delta
J_\pm\right]+J_\pm^{-1}\delta QJ_\pm.
$$
Solving this equation we get
$$
\delta J_\pm=J_\pm E(\xi)\left(\int_{\xi_0}^\xi{\rm
d}\xi'E^{-1}(\xi')J_\pm^{-1}\delta QJ_\pm
E(\xi')\right)E^{-1}(\xi)
$$
with $\xi_0=0$ for $\delta J_-$ and $\xi_0=\infty$ for $\delta
J_+$. Recalling the definition of the scattering matrix (\ref{6})
and using the relations (\ref{10}), we obtain its variation:
\begin{equation}
\frac{\delta S}{\delta\tau}=G_+\Upsilon_+(k)G_-^{-1} = H_+^{-1}\Upsilon_-(k)H_-,
\label{34}\end{equation}
where
$\Upsilon_\pm(k)\equiv\Upsilon_\pm(0,\infty;k)$ and
$$
\Upsilon_\pm (q,p;k)=\int_q^p{\rm d}\xi E^{-1}\Psi_\pm^{-1}R\Psi_\pm E.
$$
By variation of the relations (\ref{10}) and using
the above formulae one easily derives the perturbation-induced evolution of the
matrices $\Psi_\pm$:
\begin{equation}
\frac{\delta\Psi_\pm}{\delta\tau}=\Psi_\pm E\Pi_\pm E^{-1},
\label{36}\end{equation} where the  functional $\Pi$ is defined
as follows (omitting the explicit $\tau$-dependence)
\begin{equation}
\Pi_+(\xi,k) = \left(\begin{array}{cc}
\Upsilon_{+11}(0,\xi;k)&-\Upsilon_{+12}(\xi,\infty;k)\\
\Upsilon_{+21}(0,\xi;k)&-\Upsilon_{+11}(0,\xi;k)\end{array}\right),\;
\Pi_-(\xi,k) = \left(\begin{array}{cc}\Upsilon_{-11}(0,\xi;k)&\Upsilon_{-12}(0,\xi;k)\\
-\Upsilon_{-21}(\xi,\infty;k)&-\Upsilon_{-11}(0,\xi;k)\end{array}\right).
\label{37}\end{equation}
The r.h.s. of (\ref{36}) should be added to the second 
(evolutional) Lax equation (\ref{second})
to account for the perturbation-induced evolution of $\Psi_\pm$.
Note that ${\rm tr}\Pi_\pm=0$ in agreement with the 
identity ${\rm det}\Psi_\pm=1$
and $\Pi_+^\dagger(\bar k)=-\Pi_-(k)$ due to the involution (\ref{invol2}).

Formula (\ref{36}) accounts for the perturbation {\it exactly}.
However direct application of (\ref{36})  is not possible due to the fact
that $\Psi_\pm$ should be obtained from the RH problem, solution
of which requires knowledge of the spectral data, while the
perturbation-induced evolution of the latter explicitly depends
on $\Psi_\pm$ (see below). On the other hand, if the perturbation
is small, one can expand the exact
evolution equations for the spectral data into series of approximate equations which are solvable.  \\

\subsection{Evolution of the continuous datum}
In view of the boundary values of $\Psi$ (\ref{12}),
 the perturbation-induced evolution (variation) of $\beta(k)$ 
obtains from (\ref{37}) taken at the boundary $\xi=0$. We have
$$
\left(\begin{array}{cc}0&\delta\beta/\delta\tau\\
0&0\end{array}\right)=\left(\begin{array}{cc}1&\beta\\0&1\end{array}\right)
\left(\begin{array}{cc}0&-\Upsilon_{+12}(k)\\
0&0\end{array}\right).
$$
Hence the variation of $\beta(k)$ satisfies
\begin{equation}
\frac{\delta\beta}{\delta\tau}=-\Upsilon_{+12}(k)=-\int_0^\infty{\rm
d}\xi e^{2ik\xi}\left(\Psi_+^{-1}R\Psi_+\right)_{12}.
\label{38}\end{equation} Substitution of the expression for $R$
from (\ref{33}) gives
\begin{equation}
\frac{\delta\beta}{\delta\tau}=\int_0^\infty{\rm d}\xi
e^{2ik\xi}\left[\bar
r\left(\Psi_{+12}\right)^2+r\left(\Psi_{+22}\right)^2\right].
\label{39}\end{equation}
The integrand in (\ref{39}) is meromorphic for ${\rm Im}k\ge0$
by the definition of $\Psi_\pm(k)$  (\ref{11}) and this is in 
agreement of the analytic properties of $\beta(k)$.

Equation (\ref{39}) allows us to prove the applicability of the linear
limit in accounting for the effect of non-zero dephasing on the 
spike formation
in the SRS equations~\cite{24}.  Moreover, the argument we present below 
is valid for {\it non-small} dephasing parameter $\Gamma$. 
Following the arguments of 
Ref.~\cite{24} (see also Ref. \cite{14}) we note that  the  spike of 
pump radiation, first observed
in Ref.~\cite{18} (see also more recent Refs.~\cite{add1} and \cite{add2}),
occurs at a time $\tau_0\equiv t_0$ when $\beta(k_0,\tau_0)=0$. Hence,
the limit $\beta\ll1$ serves as an approximation when dealing with the spike
formation. For small $\beta$, solving the RH problem (\ref{19}) 
in the linear limit
and using (\ref{13}) we get  (in the SRS notations (\ref{2}))
\begin{equation}
u=-\frac{1}{\pi}\int{\rm d}ke^{-2ikx}\beta(k)+{\cal O}(\beta^2).
\label{linear}\end{equation}
Here the integration is in the complex plane above all poles of $\beta(k)$.
On the other hand, substitution of $\delta u/\delta \tau = - \Gamma u$ 
in equation
(\ref{39}) gives
\begin{equation}
\frac{\delta\beta}{\delta t}=\int_0^\infty{\rm
d}xe^{2ikx}\frac{\delta u}{\delta t}+{\cal
O}(\beta^3)
= -\Gamma\int_0^\infty{\rm d}xe^{2ikx}u+{\cal
O}(\beta^3)
=-\Gamma\beta+{\cal O}(\beta^3).
\label{perturb}\end{equation}
Here we have used formula (\ref{linear}).

Equation (\ref{perturb}) is a quite remarkable result. Indeed, the 
integrable part of the evolution of $\beta(k,\tau)$ 
is given by the  Riccati equation (\ref{29}), 
which is {\it quadratic} in $\beta$.
When $\beta$ is small (spike formation), addition of the 
perturbation-induced evolution
in the same order in $\beta$ to the integrable limit  requires 
keeping the terms linear and  quadratic in $\beta$. 
Howerver, in equation (\ref{perturb})
the quadratic term is equal to zero. Thereby, the  only effect of the
dephasing term is {\it stabilization} of  the spike of pump radiation 
in the SRS as it was realized, though without a rigorous argument, 
in Refs. \cite{24}.

\subsection{Evolution of the discrete data}

Here we describe the action of a perturbation on the evolution of
the discrete RH data. A perturbation-induced evolution of
$c_j^{(0)}$ is calculated from (\ref{32}) and (\ref{39}). Namely,
\begin{equation}
\frac{\delta c_j^{(0)}}{\delta\tau}={\rm
Res}\left\{\int_0^\infty{\rm d}\xi e^{2ik\xi}\left[\bar
r\left(\Psi_{+12}\right)^2+r\left(\Psi_{+22}\right)^2\right],k_j\right\}.
\label{41}\end{equation}
Separating out the singular parts of the functions 
inside the integrand by using
formula (\ref{22}),
\begin{eqnarray}
\Psi_{+12}(k)&=&\frac{c_j^{(0)}e^{-2ik_j\xi}}{k-k_j}
 \Psi_{+11}(k_j)+ \Psi_{+12}^{\rm reg}(k),\nonumber\\
& &\label{reprs}\\
\Psi_{+22}(k)&=&\frac{c_j^{(0)}e^{-2ik_j\xi}}{k-k_j}
 \Psi_{+21}(k_j)+ \Psi_{+22}^{\rm reg}(k),\nonumber
\end{eqnarray}
we can write equation (\ref{41}) as follows
\begin{eqnarray}
\frac{\delta c_j^{(0)}}{\delta\tau}&=&{\rm
Res}\Bigg[\frac{c_j^{(0)2}}{(k-k_j^2)}\int_0^\infty{\rm d}\xi
e^{2ik\xi-4ik_j\xi}F_1(\xi)\nonumber\\
&+&2\frac{c_j^{(0)}}{k-k_j}\int_0^\infty{\rm d}\xi
e^{2ik\xi-2ik_j\xi}F_2(k,\xi),k_j\Bigg]. \label{42}
\end{eqnarray}
Here we have introduced a $k$-independent, $F_1(\xi)$, and a regular
function of $k$, $F_2(\xi,k)$, by (omitting the explicit $\tau$-dependence for
simplicity)
\[
F_1(\xi)=\bar
r(\xi)\left(\Psi_{+11}(\xi,k_j)\right)^2+r(\xi)
\left(\Psi_{+21}(\xi,k_j)\right)^2,
\]
\[
F_2(k,\xi)=\bar r(\xi)\Psi_{+11}(\xi,k_j)\Psi_{+12}^{\rm
reg}(\xi,k)+r(\xi)\Psi_{+21}(\xi,k_j)\Psi_{+22}^{\rm reg}(\xi,k).
\]
Calculating the residues in (\ref{42}), we obtain the
perturbation-induced evolution of the coefficient $c_j^{(0)}$:
\begin{equation}
\frac{\delta c_j^{(0)}}{\delta\tau}=2c_j^{(0)}\left(
\int_0^\infty{\rm d}\xi F_2(k_j,\xi) +
ic_j^{(0)}\int_0^\infty{\rm d}\xi  \xi e^{-2ik_j\xi}F_1(\xi) \right).
\label{coeff}\end{equation}

Finally, the perturbation-induced evolution of $k_j$ is found as follows.
We start with the formula similar to (\ref{kj}) but with the variational derivatives, i. e.,
\begin{equation}
\left. \frac{\delta k_j}{\delta \tau}
= -\frac{ (\delta/\delta \tau)\beta^{-1}(k)}{(\partial/\partial k)\beta^{-1}(k)}\right|_{k=k_j}.
\label{variat}\end{equation}
Using  the relations
\[
\beta^{-1}(k) = \frac{k-k_j}{c_j^{(0)}}+ {\cal O}[(k-k_j)^2],\quad
\frac{\partial \beta^{-1}}{\partial k}(k) = \frac{1}{c_j^{(0)}} +{\cal O}(k-k_j),
\quad k\to k_j,
\]
which follow  from  (\ref{32}), and taking the
perturbation-induced evolution of $\beta$ in the form of
(\ref{38}),  we rewrite  (\ref{variat}) as
\begin{equation}
\left. \frac{\delta k_j}{\delta \tau}= - \frac{(k-k_j)^2\Upsilon_{+12}(k)(1+{\cal O}(k-k_j))}
{c^{(0)}_j +{\cal O}(k-k_j)}\right|_{k=k_j}.
\label{step1}\end{equation}
Note that the numerator is non-zero because $\Upsilon_{+12}(k)$ is
meromorphic function having poles of the second order, as it is seen from (\ref{reprs}).
Using (\ref{39}) and (\ref{reprs}), we obtain
\[
\Upsilon_{+12} = -\frac{c_j^{(0)2}}{(k-k_j)^2}\int_0^\infty{\rm
d}\xi e^{-2ik_j\xi}\left[\bar
r\left(\Psi_{+11}(k_j)\right)^2+r\left(\Psi_{+21}(k_j)\right)^2\right]+ {\cal O}[(k-k_j)^{-1}],
\quad k\to k_j.
\]
 Substituting this result into (\ref{step1}), we arrive at
the perturbation-induced evolution of  $k_j$:
\begin{equation}
\frac{\delta k_j}{\delta \tau}= c^{(0)}_j\int_0^\infty{\rm
d}\xi e^{-2ik_j\xi}\left[\bar r\left(\Psi_{+11}(k_j)\right)^2
+ r\left(\Psi_{+21}(k_j)\right)^2\right].
\label{last}\end{equation}

We conclude this section with the note that the general 
(i. e., with account for the perturbation)
evolution equations for the spectral data $\beta(\tau,k)$, 
$k_j(\tau)$, and  $c^{(0)}_j(\tau)$ are
given by addition of their perturbation-induced  evolution 
(represented by the variational derivatives
(\ref{39}), (\ref{coeff}), and (\ref{last})  
to the evolution equations of these quantities in the integrable limit.

\section{Conclusion}
We have shown that the Riemann-Hilbert problem is a natural
setting for dealing with both integrable and nearly integrable
nonlinear equations with the singular dispersion relations.
In particular, for small perturbations of the integrable models, we have explicit
perturbation-induced evolution equations for the RH data, which are necessary for
analysis and approximate solution for the physically interesting quantities.
\bigskip

\section*{Acknowledgments}
We are grateful to Professors J.-G. Caputo and P. Sabatier for
their support which made our participation in the R.C.P. 264
conference (Montpellier, 2000) possible. We are also indebted to
Professor J. Leon for useful comments and Doctor A. Grabtchikov
for discussions about experimental aspects of SRS. The work by
E.D. was supported by the grant No. 97-2018 from INTAS-Belarus.

\end{document}